# The genetic code multiplet structure, in one number

Tidjani Négadi
Physics Department, Faculty of Sciences,
Oran University, Oran, 31100, Algeria
physicants@aol.com[1]

**Abstract**. The standard genetic code multiplet structure as well as the correct degeneracies, class by class, are all extracted from the (unique) number 23!, the order of the permutation group of 23 objects.

## 1. Introduction.

In this short paper, we report briefly on a striking observation made by us (January 2007) in the course of an investigation aiming at using a novel group theoretical classification scheme of the 20 amino acids, based on the use of the Symmetric Group $S_{23}$, or permutation group of 23 objects [2]. The genetic code multiplet structure as well as its various correct numeric degeneracies seem literally *encrypted as* (and *in*) the digits of the factorial of the number twenty-three, the order of $S_{23}$. The relevance of 23 as a "magical number" has been emphasized thirty seven years ago by Gavaudan [1]. He showed in particular that the existence of 23 types of signals (20 amino acids *and* 3 stops) is the price to code for 20 amino acids and, as a consequence, he established several (mathematically remarkable) numerical ratios involving 23 and other important numbers like 20, the number of amino acids, 44, the total degeneracy including the 3 termination signals or stops or 64, the total number of possible codons.

The today experimentally well established 64 codons multiplet structure of the *standard* genetic code, used by the great majority of living beings on Earth, is given by the following pattern

- 9 doublets: {C, N, D, K, Q, E, H, F, Y}
- 5 quartets: {G, A, P, V, T}
- 3 sextets: {S, L, R}
- 2 singlets: {M, W}   (P$_1$)
- 1 triplet: {I}
- 3 stops: {Amber, Ochre, Opal}

where the amino acids are written in the one-letter code. At this point, one might recall the memorable experimental work done by Nirenberg and Khorana, which began in 1961 (Nirenberg and Matthaei [3]) and led in only some few years to the emergence of the genetic code table. Recall that there exist 64 ($4^3$) triplet-codons, of which 61 are *meaningful* and code for 20 amino acids and 3 are termination signals or stops. There are therefore 41 degenerate codons but this last number rises to 44 (=3+41) if the stops are included. This is for the standard genetic code with the multiplet structure shown above. Another important well known view of the 20 amino acids to mention here, because it nicely manifests itself in this work, singles out the two amino acids methionine M and tryptophane W from the rest (18 amino acids). They (M and W) are both coded by only one codon and therefore have degeneracy 0. Correspondingly, there are 59 codons for which there exist a *synonymous alternative* (61-2).

---
[1] Alternate email : negadi_tidjani@univ-oran.dz

In a first section we establish the characteristic numbers of the standard genetic code directly from the number 23! (also the order of the symmetric group $S_{23}$) and show that they are literally *encrypted* in its mathematical features. The gross features of the degeneracy are obtained. In a second section, we establish the multiplet structure of the genetic code as well as the detailed degeneracy for each one of the 5 classes, from the digits themselves. Finally, section 4 contains some other interesting elements, having to do with the physico-chemical characteristics of the amino acids (nucleons, carbon, oxygen, nitrogen and sulphur atoms) and the pattern of their distribution.

## 2. 23! and the standard genetic code degeneracy

There are two ways to look at the degeneracy. First the usual one, which is related to the third base in a codon. In this case one has, as the degeneracy pattern, the above (experimental) classification {9,5,3,2,1} where the 3 sextets have each 6 codons (4+2) among which 5 are degenerate, so that the total degeneracy is $9\times1+5\times3+3\times5+2\times0+1\times2=41$ and the addition of 20 amino acids (codons) gives 20+41=61. The other case, is related to the degeneracy at the *first* base in the codon. Here, 17 amino acids are "non-degenerate" and 3 are doubly "degenerate" $S^{II}$, $S^{IV}$, $R^{II}$, $R^{IV}$, $L^{II}$ and $L^{IV}$ with pattern 17+6=23. Considering $X^{II}$ and $X^{IV}$ (X=S, R, L) as distinct entities, we have now for the total degeneracy $9\times1+5\times3+\underline{3\times(1+3)}+2\times0+1\times2=38$, and therefore $(17+2\times3)+38=23+38=61$.

Now consider 23! (also equal to the total number of permutations in the symmetric group $S_{23}$ or its order)

$$23!=25852016738884976640000 \qquad (1)$$

This number is stupendously interesting as its has 23 digits, in base-10. (23 is the unique *odd* number the factorial of which has this property.) There are 18 non vanishing digits and 5 zeros so that there is a perfect matching with the corresponding numbers for the twenty amino acids and the three termination signals (or stops): 18 amino acids with *degenerate* associated codons, 2 amino acids with *degeneracy*-0 codons and 3 stops to which the value 0 is usually assigned. This striking mathematical fact will be pushed farther below. The decimal expression of the factorial in equation (1), in base-100, writes

$$23!_{100} = [2,58,52,1,67,38,88,49,76,64,0,0]$$
$$= 25852167388849766400 \qquad (2)$$

and we have explicitely shown its 12 "digits" (the same digital symbols as those in Eq.(1), most of them grouped in pairs). The point is that, when $23!_{100}$ *is* *interpreted* as a base-10 number, then the one-symbol digits separate and there is now 18 non vanishing digits and 2 zeros, just the right numbers for the 20 amino acids, and the 3 stops (3 zeros) *disappeared* in the process. Let us consider also the prime decomposition of 23!, using the fundamental theorem of number theory

$$23! = 2^{19} \times 3^9 \times 5^4 \times 7^3 \times 11^2 \times 13 \times 17 \times 19 \times 23 \qquad (3)$$

Aside the fact that this latter form is also informative, as the above decimal representation, it possesses also a "multiplicity" structure which is seen in the exponents: the prime numbers (the building blocks of the numbers) could appear more than one time, in analogy with the codons degeneracy. The "morphological" characteristics of the numbers in eq.(1) are the following

$$\eta = 23 \qquad (4)$$
$$\eta_0 = 23 - 5 = 18 \qquad (5)$$
$$\eta^0 = 20 \qquad (6)$$

Here, $\eta$ is the total number of digits (including zeros and multiplicities), $\eta_0 = \eta - 5$ the number of non vanishing digits and $\eta^0$ (=20) the total number of digits in $23!_{100}$ (see above). The *arithmetical* characteristics, from Eq.(3), are the following

$$\varepsilon = 41 \qquad (7)$$
$$\pi = 100 \qquad (8)$$
$$\rho = 200 \qquad (9)$$

where $\varepsilon$ is the sum of the exponents $\varepsilon_i$, $\pi$ the sum of the prime factors $\pi_i$ and $\rho$ the sum of the products $\varepsilon_i \times \pi_i$. It is striking that Eq.(7), $\varepsilon = 41$, which is the sum of the exponents or, as we mentioned above, the sum of the multiplicities of the prime factors coincides with the sum of the degeneracies of the 20 amino acids. (In fact, it could do more as we shall show below; it is also related to proline's singularity and coincides with the number of nucleons in its side-chain.) Considering $\varepsilon$ as representing the total degeneracy and adding it to each of the numbers in Eqs.(4)-(6), we have

$$\eta + \varepsilon = 23 + 41 = 64 \qquad (10)$$
$$\eta_0 + \varepsilon = 18 + 41 = 59 \qquad (11)$$
$$\eta^0 + \varepsilon = 20 + 41 = 61 \qquad (12)$$

These three equations sum up our results in this section, all infered from the number 23!. Eq.(10): 23 signals (20 amino acids and 3 stops) added to 41 (the total number of degenerate codon) gives the correct total number of codons, Eq.(11): it reconstitutes the correct number of codons for which there exist a synonymous alternative (see the introduction) and finally Eq.(12): it gives the number of meaningful codons. Note, as a remark, that $\pi - \varepsilon = 59$.

Other interesting relations could be infered from Eqs.(7)-(9) only; we give in this paper only some ones. Consider the sum

$$\Xi := \varepsilon + \pi = 141 \qquad (13)$$

It has the same looking in base-100 ([1,41]) except that the digits are "1" and "41" which is interesting as it reminds us proline's singularity (see the remark above). As it is well known, proline (a secondary amine acid or less rigorously an imino acid) has no well definite side-chain contrary to the other 19 amino acids. Shcherbak [4] have found some years ago a clear manifestation of decimal arithmetic inside the genetic code and published remarkable (and beautiful) nucleon number balances, at the price of the following mathematical trick consisting in "borrowing" *one* nucleon from the side chain with 42 nucleons to the block which is the same, 74 nucleons, for 19 amino acids but only 73 for proline. Taking the sum of digits (in base-100), we obtain **SOD**($\Xi$)=1+41=42. Next, we consider the sum

$$\Sigma := \varepsilon + \pi + \rho = 341 \tag{14}$$

We have immediately **SOD**($\Sigma$)=3+41=44. Considering our original interpretation of the number 41 as the total number of degenerate codons, we recover the 3 stops added to give the characteristic number 44. As a third and last example here, we take all equations (4)-(9) together and compute their mean

$$(\varepsilon + \pi + \rho + \eta + \eta_0 + \eta^0)/6 = 67 \tag{15}$$

This is nothing but the total number of carbon atoms in the 20 amino acids. In the fourth section this number, and others, will be recovered from the digits themselves.

## 3. The amino acids as the decimal digits

Here we show that simple assumptions, driven by logic could lead us to establish a strong link between the 23 digits, on the one hand, and the 20 amino acids and the 3 stops, on the other, and, as a bonus, an easy computation of the degeneracy of the detailed multiplet structure of the (standard) genetic code given in the introduction, completing in this way the results obtained in the preceeding section. Consider collectively the set of 23 digits of 23! in Eq.(1)

$$\{0,0,0,0,0,1,2,2,3,4,4,5,5,6,6,6,7,7,8,8,8,8,9\} \tag{16}$$

Here, the digits are "laid bare" of their vital quality, i.e., their *weights*, but in Eq.(1) these latter are apparent and effective. In the above set, *all* the 10 symbols of the decimal system are present (0,1,2,…,9), some with multiplicity. A clear partition of the integers consist of an even sub-set of the form 2n and a odd sub-set of the form 2n+1. The primes constitute a highly important sub-set of the latter but with no simple formula (at least at the present time) as they are the building-blocks of the numbers. In the spirit of this (natural) partition, let us, first, *exclude the zeros* and choose as a sorting criterion into sub-sets the following first claim by using the logical connectives **and** and **not**:

$$\text{Even } \textbf{and not}(\text{Prime}) \tag{$C_1$}$$

This selects 9 even numbers which are not primes, 4 two times, 6 three times and 8 four times (2 is excluded). This sub-set is quite interesting as there are precisely 9 doublets. For the other remaining sub-set with numbers obeying **not**($C_1$) or explicitely Odd **or** Prime, by DeMorgan's Laws, our second claim is

$$\text{Odd } \textbf{and} \text{ Prime} \tag{$C_2$}$$

This gives 5 odd primes 3 one time, 5 two times and 7 two times and so we guess the 5 quartets. There remains the set {1,2,2,9} to be partitioned, to attribute the last digits, and we have found interesting to take {1,2,9} for the 3 sextets and {2} for the sole triplet. This choice will appear correct below (see also the last words at the end of this paper). We sum up the above sorting as the following (digit) pattern, to be compared to the one in the introduction:

- 9 doublets: {4,4,6,6,6,8,8,8,8}
- 5 quartets: {3,5,5,7,7}
- 3 sextets:  {1,2,9}
- 2 singlets: {0,0}                              (P$_2$)
- 1 triplet:  {2}
- 3 stops:    {0,0,0}

Call $\nu_d$ the *number of digits* and $\sigma_d$ the *sum of digits* (*without repetition*), in a given set; the subscript d indicating the degeneracy-class number (d=1,2,3,4,6). We show below that $\nu$, $\sigma$ and their sum contain all what is necessary to compute the degeneracies. We obtain

- Doublets: $\nu_2 = 9$, $\sigma_2 = 18$, $\nu_2+\sigma_2 = 27$
- Quartets: $\nu_4 = 5$, $\sigma_4 = 15$, $\nu_4+\sigma_4 = 20$
- Sextets:  $\nu_6 = 3$, $\sigma_6 = 12$, $\nu_6+\sigma_6 = 15$
- Singlets: $\nu_1 = 2$, $\sigma_1 = 0$, $\nu_1+\sigma_1 = 2$       (P'$_2$)
- Triplet:  $\nu_3 = 1$, $\sigma_3 = 2$, $\nu_3+\sigma_3 = 3$
- Stops:    $\nu_{stops} = 3$, $\sigma_{stops} = 0$, $\nu_{stop}+\sigma_{stop} = 3$

In a first analysis of theses numbers, we see that the degeneracy is encoded differently for the quartets, singlets and the triplet, on the one hand, and the doublets and the sextets, on the other. For the former case, in *each* degeneracy class, the number of amino acids is given by $\nu_d$, the number of degenerate codons by $\sigma_d$ and the number of total codons by their sum $\nu_d+\sigma_d$. For the latter case, the number of amino acids is given by $\nu_d$, a property shared by all the five classes, but for degeneracies things appear somewhat "shifted". In this case, we would have

- Doublets: $\sigma_2-\nu_2=9$ gives the number of degenerate codons and $\sigma_2=18$ total number of codons.
- Sextets: $\sigma_6=12$ gives the number of degenerate codons, 3×(3+1), in the sense mentioned at the begining of the preceeding section, $\nu_6+\sigma_6 = 15$ gives the number of degenerate codons, in the usual sense 3×5.

This seem quite nice but we can do better, thanks to the basic symmetry of the genetic code. As a matter of fact Findley and his collaborators [5] in a well known group theoretical study of the genetic code viewed this latter as a *relation* and this led them to extract its inherent degeneracy-2 basic symmetry by arranging the 64 codons table according to the third base in the codons. Rephrasing this symmetry in terms of a logical double implication they established, for the degeneracy-2, the existence of a *one-to-one* correspondence, of *exact nature* [5], from one member of a doubly degenerate codon pair to the other. This basic symmetry seems beautifully realized in our pattern (P$_2$). First, all the digits are even and, second, there are 6 digits in the form 2(2n) and 3 digits in the form 2(2n+1). This is in excellent agreememt with the fact that there are 6 cases with exclusively *pyrimidine* third bases (F, Y, C, H, N, D) and 3 cases with exclusively *purine* third bases (Q, K, E). It is now clear that the existence of the one-to-one correspondence mentioned above allows us to *halve* $\sigma_2$ for the doublets and take instead $\sigma'_2 = 9$. Armed with this value, we could now compute easily the detailed degeneracies. First, by adding all the σ- and ν- contributions (including the stops) we have

$$(\nu_1+\nu_2+\nu_3+\nu_4+\nu_6) + \nu_{stops} = (2+9+1+5+3)+3 = 23$$

$$(\sigma_1+\sigma'_2+\sigma_3+\sigma_4+\sigma_6)+\sigma_{stops} = (0+9+2+15+12)+0 = 38$$

(17)

and their sum gives 23+38=61. This number is just the number of meaningful codons, with the 3 stops missing! A way out of this dilemma is to let the 3 sextets and the 3 stops have some kind of "interaction" and re-write the first equation (17) as $(\nu_1+\nu_2+\nu_3+\nu_4)+(\nu_6+\nu_{stops})$ or 17+(3+3)=23. This latter re-writing together with the second equation (17) define exactly the first-base degeneracy structure, mentioned at the beginning of section 1, at the sole condition to make the identification (3+3) ↔ ($S^{II}$, $R^{II}$, $L^{II}$)+($S^{IV}$, $R^{IV}$, $L^{IV}$). This "entanglement" involving the 3 stops and the 3 sextets could, may-be, find a support in the following observation concerning the use, by the (known) variant forms of the genetic code deposited at NCBI [6], of "non-standard" termination codons. As a matter of fact, besides the usual stop codons (Amber, Ochre and Opal), we have found the following 5 only cases (out of 17) all associated to a mitochondrial genetic code where the stop-sextet relation is clear (experimentally established): (i) *Vertebrate mitochondrial code*, (ii) *Ascidian mitochondrial code*, (iii) *Chlorophycean mitochondrial code*, (iv) *Scendesmus obliquus mitochondrial code*), (v) *Thraustochytrium mitochondrial code*. (Also known [7] is the case of the Spiroplasma and Mycoplasma molicutes where the codon CGG normally coding for arginine becomes a termination signal.) Second, by proceeding as above but this time counting the sextets *two times*, as it is required by, first, symmetry considerations (Rumer's symmetry) and, second, as it insures the existence of many physico-mathematical regularities discovered by Shcherbak [4] (see below). We could write either

$$(\nu_1+\nu_2+\nu_3+\nu_4+\nu_6) + \nu_{stops} = (2+9+1+5+3)+3 = 20+3=23$$

$$\sigma_1+\sigma'_2+\sigma_3+\sigma_4 + (\sigma_6+\nu_6) = 0+9+2+15+15 = 41$$

(18)

$$20+41+3=64$$

in the usual view, or

$$\nu_1+\nu_2+\nu_3+\nu_4+2\nu_6 = 2+9+1+5+2\times3 = 17+6=23$$

$$(\sigma_1+\sigma'_2+\sigma_3+\sigma_4+\sigma_6)+\nu_{stops} = (0+9+2+15+12)+3 = 38+3=41$$

(19)

$$23+38+3=64$$

in the "first-base degeneracy" view considered in this paper. The parts of Eq.(18) and Eq.(19) are meant added. In both views, the correct number of codons and the correct degeneracies, for each one of the five degeneracy classes, are obtained as it could be easily read from Eq.(18) and Eq.(19), just by alining the corresponding $\sigma_i$'s and $\nu_i$'s (i=1,2,3,4,6).

## 4. Miscellaneous

In this last section, some few remarkable relations, expressing a (mysterious) link between the total carbon atom number and the total nucleon number, on the one hand, and the total number of carbon, oxygen, nitrogen and sulfur atoms (CNOS) and the total number of atoms, on the other, are derived from the physico-chemical characteristics of the 20 amino acids, using only some elementary arithmetic operations. In the table below, we give (i) the number of carbon atoms, (ii) the number of carbon, oxygen, nitrogen and sulfur atoms (CNOS), (iii) the number of atoms and finally (iv) the number of nucleons, or integer molecular weight. (All the numbers in the table refer to the side-chains except for the number of atoms where the blocks are also included; d is the degeneracy class number.)  From the table, we have 67 carbon atoms in the 20 amino acids and 76 carbon atoms, *counting the sextets two times*. There are 87 CNOS atoms in the 20 amino acids and 100 CNOS atoms, *counting the sextets two times*, see [8]. Finally, there are 384 atoms (204 in the side-chains) and 1255 nucleons, in the 20 amino acids side-chains (1443 nucleons in the side-chains and 3145 in the side-chains-and-blocks, *counting the sextets two times*). First, we have from Eq.(P'$_2$) by suming over the degeneracy classes (i=1,2,3,4,6 and the Findley-Findley-McGlynn (FFMcG) degeneracy-2 one-to-one correspondence is not used)

$$\Gamma = \nu_1+\sigma_1+\nu_2+\sigma_2+\nu_3+\sigma_3+\nu_4+\sigma_4+\nu_6+\sigma_6= 67 \qquad (20)$$

This coincides with the number of carbon atoms in the 20 amino acids; note that the 3 stops were excluded.. Supporting this affectation is the following fact: by partitioning the above sum into two parts, the quartets, the singlets and the triplet, on the one hand as their $\nu+\sigma$ sum gives exact results for the degeneracies (see above), and the doublets and the sextets, on the other as they do not, the corresponding numeric partition is then 25+42=67. It is significative to remark that the number of carbon atoms in the *corresponding* sub-parts conforms exactly to the *same* pattern: 9+12+4=25 and 33+9=42, respectively (see the table).

| d | aa | # carbons | # CNOS | # atoms (s.c.) | # nucleons |
|---|---|---|---|---|---|
| 4 | P | 3 | 3 | 17 (8) | 41 |
|   | A | 1 | 1 | 13 (4) | 15 |
|   | T | 2 | 3 | 17 (8) | 45 |
|   | V | 3 | 3 | 19 (10) | 43 |
| 6 | G | 0 | 0 | 10 (1) | 1 |
|   | S | 1 | 2 | 14 (5) | 31 |
|   | L | 4 | 4 | 22 (13) | 57 |
|   | R | 4 | 7 | 26 (17) | 100 |
| 2 | F | 7 | 7 | 23 (14) | 91 |
|   | Y | 7 | 8 | 24 (15) | 107 |
|   | C | 1 | 2 | 14 (5) | 47 |
|   | H | 4 | 6 | 20 (11) | 81 |
|   | Q | 3 | 5 | 20 (11) | 72 |
|   | N | 2 | 4 | 17 (8) | 58 |
|   | K | 4 | 5 | 24 (15) | 72 |
|   | D | 2 | 4 | 16 (7) | 59 |
|   | E | 3 | 5 | 19 (10) | 73 |
| 3 | I | 4 | 4 | 22 (13) | 57 |
| 1 | M | 3 | 4 | 20 (11) | 75 |
|   | W | 9 | 10 | 27 (18) | 130 |
| Total sum | | 67 | 87 | 384 (204) | 1255 |

Note also that, by using the iterated (complete) **SOD** function, we have Γ+**SOD**(23!) =67+9=76. Second, we consider the total number of nucleons in the 20 amino acids, 1255. Interestingly, this number is also the total number of partitions of the number 23 and therefore the total number of irreducible representations of the symmetric group $S_{23}$ (see the introduction). In base-100 its digits are "12" and "55", so that their sum is 12+55=67. Better, we have 12 carbon atoms in the 2 degeneracy-0 singlets (M and W) and 55 carbon atoms in the 18 remaining amino acids with non-vanishing degeneracy (see section 2 for the great importance of the partition 2+18). If we now take 3145 the total number of nucleons (including the blocks of the amino acids see [3]), then we have by proceeding as before 31+45=76 which is 67+9, see above. Third, 384 the total number of atoms in the 20 amino acids, in base-100, has as digits "3" and "84", so that 3+84=87, i.e., just the total CNOS number, in the 20 amino acids. Here, we surmise some manifestation of the partition 1+19 (proline+19 amino acids) because, first, proline seems to have some importance (see section 3) and, second, it has only 3 carbon atoms in its side-chain (no oxygen, no nitrogen and no sulfur atoms) while there are 84 atoms in the remaining 19 amino acids. As a final remark, having to do with the well known fact that the degeneracy at the third base of a codon is correlated with the molecular mass (here the number of nucleons), let us consider, in each degeneracy class, the difference between the highest nucleon number and the lowest one; call it $\Delta\mu_d$. Using the table we have $\Delta\mu_1$=55, $\Delta\mu_2$=60, $\Delta\mu_3$=0 (as isoleucine is unique), $\Delta\mu_4$=44 and $\Delta\mu_6$=69. First, by taking the same sub-sets leading to the pattern accompanying Eq.(20), mentioned above (42+25, digits-and-carbon number), we have $\Delta\mu_2+\Delta\mu_6$=129 and $\Delta\mu_1+\Delta\mu_3+\Delta\mu_4$=99. Taking a simple sum of digits for the former, that is 1+2+9, gives a two-fold interesting result: (i) it looks like "fitting" the degeneracy structure of the 3 sextets L (CUN+UUR), R (CGN+AGR) and S (UCN+AGY). Leucine and Arginine have the same first base in their quartet parts, they have both purine third bases in their doublet parts and the same middle base in each of them (quartet+doublet) while serine has different middle bases and pyrimidine third bases. Serine is therefore aside and the degeneracy is 3 for the 3 doublet-parts, 2 for L and R (2 purine third bases) and 1 for S (1 pyrimidine third base) and 9 for the 3 quartet-parts, or (1+2)+9 and (ii) it gives (also) the correct number of amino acids in class 2 (9) and in class 6 (1+2), or 1+2+9=12. As for the latter, 99, by using a simple two-step process involving the **SOD** and also the **SOPF** (Sum Of Prime Factors) functions, it reproduces the correct total number of amino acids in classes 1, 3 and 4: **SOPF**(**SOD**(99))= **SOD**(**SOPF**(99))=8. In summary this last result together with (ii) leads to 12+8=20. Note finally that the three ordered Δμ-values of the even classes 44, 60 and 69 lead through substractions to 16+9=25 or $4^2+3^2=5^2$ which is the Pythagorean triple found by Shcherbak [4]. Second, the total contribution of the even classes gives $\Delta\mu_2+\Delta\mu_4+\Delta\mu_6$=173. The addition of its digits in base-100 describes proline's block 1+73=74 and 1+41 from Eq.(13), re-written as 42-1=41, describes its side-chain, see Shcherbak's Figure 6, [4]. Suming the two parts gives (i) 115 the total number of nucleons in proline and (ii) an easy derivation of Shcherbak's trick in the form "−1+1=0". Finally, the Δμ for the whole set of 5 classes is also equal to 129 (=130-1) and, adding 59 (=π-ε, see a remark following Eqs.(10)-(12)) which is the difference between the sum of the prime factors and the sum of the exponents of 23!) gives 188 which is nothing but the number of nucleons in serine, leucine and arginine (31+57+100). Moreover, applying the FFMcG correspondence to this latter gives 188/2=94 which has the (unique) prime decomposition 2×47, and we have precisely that serine, on the one hand, has 2 CNOS atoms and leucine and arginine, on the other, have respectively 4 and 7 CNOS atoms (see the table) and that these two digits are concatenated, or "linked", according to the pattern discussed above. It seem therefore likely that the digital choice made for the sextets (1, 2 and 9) in Eq.($P_2$) was wise.

Since the deciphering of the genetic code in the sixtees of the last century, the degeneracy problem has challenged many theorists. For example, the introduction of the concept of symmetry and the concrete use of group theoretical technics, mainly inherited from particle physics and initiated by Hornos and Hornos [9] in the ninetees (see also [10] and the references therein), helped many followers but did not solve it *completely*. Other approaches, as the one using generalized information functions [11], gave interesting results but at the price of fitting parameters. Recently, two new approaches based on the use of p-Adic Mathematics, [12], and Number Theory, [13], entered upon but none has, today, solved the problem completely. In this paper we have, first, shown a clear connection with Arithmetics and, second, discovered that the multiplet structure as well as the correct degeneracies are both encoded as-and-in the digits of one unique number, 23!, the order of the permutation group of 23 objects.

## References

[1]   Gavaudan, P. (1971) in Chemical Evolution and the Origin of Life, Eds. R. Buvet and C. Ponnamperuma, North-Holland Publishing Company.
[2]   Négadi, T. (2007) Work in preparation.
[3]   Nirenberg, M.W. and Matthaei, J.H. (1961) Proc. Natl. Acad. Sci. USA, Vol.47, 1588.
[4]   Shcherbak, V. I. (2003) Biosystems 70, 187.
[5]   Findley, G. L., Findley, A. M. and McGlynn, S. P. (1982) Proc. Natl. Acad. Sci. USA, Vol.79, 7061-7065.
[6]   Elzanowski, A. and Ostell, J. http://130.14.29.110/Taxonomy/Utils/wprintgc.cgi?mode=c (2000).
[7]   Melcher, U. (1997) http://opbs.okstate.edu/~melcher/MG/MGW4/MG431.html
[8]   Yang, C.M. (2004) Bull. Math. Biol., 66(5), 1241; Journal of Biological Systems, special issueVol.12, 21.
[9]   Hornos, J.E.M. and Hornos, Y.M.M., (1993) Phys. Rev. Lett., 71, 4401-4404.
[10] Hornos, J.E.M., Braggion, L. and Magini M., (2001) Symmetry: Culture and Science, 12, Nos. 3-4, 349-369.
[11] Alvager, T., Graham, G., Hutchison, D. and Westgard, J. (1994) J. Chem. Inf. Comput. Sci., 34, 820-821.
[12] Dragovich, B. and Dragovich, A. http://arxiv.org: q-bio/07070764 v1 (2007)
[13] Khrennikov, A. and Nilsson, M., http://arxiv.org: q-bio.OT/0612022 (2006)